## 2D metals
# Flat and safe under the graphene sheet
Claire Berger and Walt A. de Heer

Large-scale atomically thin metals can be stabilized through confinement epitaxy at graphene / SiC interface, which exhibit a gradient bonding type and are air stable, providing a compelling platform for quantum and optoelectronic technologies.

As this year marks the 40th anniversary of the discovery of the Quantum Hall Effect (QHE)[1], we are reminded that matter in low dimensions may behave in strange ways. Zero resistance and quantized resistance plateaus in the QHE regime, as well as persistence of or appearance of superconductivity in atomically thin films (such as monolayer Pb, $NbSe_2$, Tin or bilayer graphene) are just a few examples. At the time of the QHE discovery, tricks were applied to confine electrons electrostatically at the interface between two materials. Today, dimensionality effects can be studied in materials that are as two-dimensional (2D) as they can get, that is one-atom thick sheets. These 2D materials can be prepared at the interfaces of two crystals, or by deposition on the surface of a substrate, or from naturally layered materials like graphite where in plane chemical bonds are much stronger than the out of plane bonds allowing easy separation of atomically flat layers[2]. In contrast to layered materials, freestanding ultrathin metal films are unstable: they will spontaneously crumple up and require a substrate to stabilize them. In addition, they are in general extremely sensitive to air and can only be studied using in-situ ultrahigh vacuum methods. Now published in Nature Materials, Natalie Briggs and coworkers[3] have found a way to easily produce 2D crystalline metals (notably non-noble metals) that can be studied in ambient conditions.

The atomically thin metal layer is sandwiched between a substrate and an over-layer, which provides protection from environmental degradation. But rather than growing a metal film on a substrate and then covering it with a protecting coating, here the metal atoms were made to spontaneously squeeze themselves under an epitaxial graphene (epigraphene) layer – a graphene that was first grown on a single crystal silicon carbide (SiC) substrate[4, 5]. Epigraphene is known to cover continuously and seamlessly the entire SiC wafer[5]. In order to promote metal atoms intercalation the graphene layer was first exposed to oxygen plasma to introduce holes, through which the metal atoms could pass, as shown in Figure 1. These defects obviously compromise the very protection that epigraphene was to provide. But the high density of defects in graphene apparently *healed* during the intercalation process at elevated temperatures (≈700 ℃). Using this technique encapsulated ultra-thin films of one to three atomic layers of Ga, In and Sn were produced[3], realizing what the authors called 'half van der Waals metals' where the atoms are bonded covalently to the bottom SiC surface thereby stabilizing the 2D layer, and at the same time are weakly bonded to the top graphene layer protecting them from the environment. By virtue of the atomic alignment between the crystal lattice of the SiC and that of 2D metals (hetero-epitaxy), the atomic structure of the

SiC/metal/graphene heterostructure stack is the same everywhere, which can be reproducibly scaled to the SiC substrate wafer size (commercially available up to 15 cm in diameter).

These 2D metal films are fully crystallized, but even more interestingly, their atomic structure, having a half-covalent (with SiC) and half van der Waals (with graphene) - bonded characteristic, is distinct from their natural 3D counterpart, giving rise to intriguing properties. For example, a zero resistance critical superconducting temperature of 3.2K for a 3-layer Ga film was measured, that is three times that of bulk Ga (1.08K), and half that of metastable β-Ga ($\approx$ 6K), but with much higher critical fields. Remarkably, the measurement was performed after the heterostructure had been exposed to air, demonstrating the environmental protection provided by the epigraphene sheet.

Ga, In and Sn are not the only elements that can slip through the graphene sheet to reach the SiC interface. The ease with which foreign atoms intercalate at the SiC-epigraphene layer interface has been known and various degrees of intercalation and stability have been demonstrated with Al, Au, Bi, Cu, F, Fe, $FeCl_3$, GaN, Ge, H, $H_2O$, Li, Mn, N, Na, O, Pb, Pt, Pt, Si, Yb, et cetera [5]. Stabilizing and protecting 2D thin films by intercalation (or *confinement heteroepitaxy*[3] in the present case) can in principle be widely applied to provide a platform for mediating compelling structural and physical properties. It will be very informative to study other metal heteroepitaxial monolayers. For instance it is recently shown[6] that continuous gold films fabricated at the epigraphene/SiC interface have unexpected semiconducting properties. Extension of this method to heavy metal Sn can help to investigate exotic pairing mechanisms in 2D superconductors[7]. In turn, intercalated layers do modify the properties of epigraphene. Separated from the substrate by hydrogen intercalation, the epigraphene layer which is previously in direct contact with SiC (buffer layer) turns from semiconducting to conducting[8]; the epigraphene layer above the buffer layer changes from heavily negatively doped to charge neutral with Sn intercalation[9], or to p-doped, controlled by the H intercalation[10]

Atom intercalation opens possibilities even beyond the stabilization of new 2D structures and the study of unknown phenomena at this scale. Epigraphene, this single or multi-layer graphene that forms epitaxially on SiC crystals when they are heated[11], has been developed as a platform for 2D nanoelectronics since 2001[4, 5], taking advantage of the perfectly defined interfaces resulting from the SiC/graphene heteroepitaxy, and the transfer-less wafer-size scalability provided by the industrial grade single crystal substrate. Having all-epitaxial 2D heterostructures that can be grown controllably and that can be studied in ambient condition is a formidable playground. One can only imagine the possibilities opened by combining epigraphene properties (room temperature ballistic conductance, record long spin diffusion, sturdy QHE plateaus, etc) with that of magnetic or superconducting 2D metals or other topological materials for investigating topological phenomena, as well as developing spintronics or advanced optoelectronics.

Claire Berger is at the UMI 2958 Centre National de la Recherche Scientifique - Georgia Institute of Technology, Metz, France and at the School of Physics, Georgia Institute of Technology, Atlanta, USA.

Email: and claire.berger@cnrs.fr and claire.berger@physics.gatech.edu

Walt A. de Heer is at the School of Physics, Georgia Institute of Technology, Atlanta, USA and the Tianjin International Center for Nanoparticles and Nanosystems, Tianjin University, Tianjin, China
Email: deheer@physics.gatech.edu

The authors declare no competing interests.

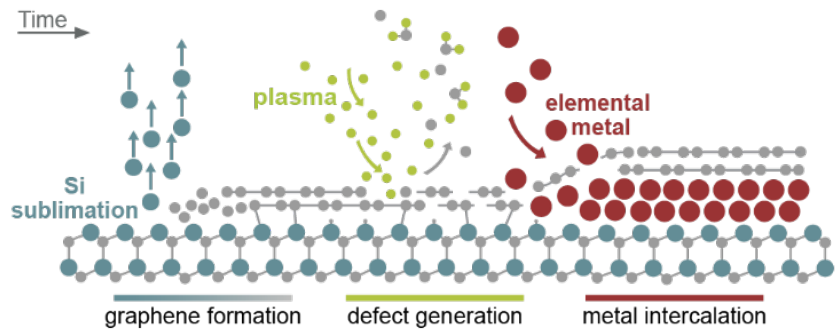

**Figure 1**. Schematic showing the process of graphene growth via Silicon sublimation, plasma treatment to generate defects, and diffusion of metal atoms through defects (intercalation), realizing an all-epitaxial epigraphene/metal/SiC structure (from[3]).